\documentclass[aps,pre,preprint,superscriptaddress]{revtex4-1}

\usepackage[pdftex]{graphicx}
\usepackage{amssymb,amsfonts,amsmath}
\usepackage{grffile}

\begin{document}

%\preprint{}

\title{Self-organized alternating chimera states in oscillatory media}



\author{Sindre W. Haugland}
\affiliation{Physik-Department, Nonequilibrium Chemical Physics, Technische Universit\"{a}t M\"{u}nchen,
  James-Franck-Str. 1, D-85748 Garching, Germany}

\author{Lennart Schmidt}
%\thanks{}
\affiliation{Physik-Department, Nonequilibrium Chemical Physics, Technische Universit\"{a}t M\"{u}nchen,
  James-Franck-Str. 1, D-85748 Garching, Germany}
\affiliation{Institute for Advanced Study - Technische Universit\"{a}t M\"{u}nchen,
  Lichtenbergstr. 2a, D-85748 Garching, Germany}

\author{Katharina Krischer}
\email[]{krischer@tum.de}
\affiliation{Physik-Department, Nonequilibrium Chemical Physics, Technische Universit\"{a}t M\"{u}nchen,
  James-Franck-Str. 1, D-85748 Garching, Germany}


\date{\today}

\begin{abstract}
Oscillatory media can exhibit the coexistence of synchronized and 
desynchronized regions, so-called chimera states, for uniform 
parameters and symmetrical coupling. In a phase-balanced chimera 
state, where the totals of synchronized and desynchronized regions, 
respectively, are of the same size, the symmetry of the system predicts that 
interchanging both phases still gives a solution to the underlying 
equations. We observe this kind of interchange as a self-emerging 
phenomenon in an oscillatory medium with nonlinear global coupling. An 
interplay between local and global couplings renders the formation of 
these alternating chimeras possible.
\end{abstract}

%\keywords{}

\maketitle

%\section{}
%\subsection{}
%\subsubsection{}

%\section{Introduction}
% The separation of a system of identical oscillators into two regions, 
% composed of mutually synchronized and desynchronized oscillators, 
% respectively, was investigated for the first time in 2002 
% \cite{Kuramoto2002} by Kuramoto \& Battogtokh, and named a ``chimera 
% state'' two years later \cite{Abrams2004} by Abrams \& Strogatz.
% % with distributed frequencies 
% Since then, the coexistence of synchrony and incoherence in oscillatory 
% systems has been the subject of several theoretical studies 
% \cite{Shima2004, Martens2010, Sethia2008, Abrams2008, Omelchenko2011, 
% Omelchenko2013, Nkomo2013, Sethia2013, Schmidt2014, Sethia2014}, and 
% chimera states have been realized in a number of different experimental 
% systems \cite{Tinsley2012, Hagerstrom2012, Nkomo2013, Martens2013, 
% Wickramasinghe2013, Schmidt2014}. 

Synchronization phenomena are omnipresent in nature, and, consequently, 
their theoretical description has received great attention for the last 
two decades \cite{Pikovsky2003}.
Given a network of coupled oscillators with a distribution of natural 
frequencies, it was a seminal achievement to describe their 
synchronization transition when the coupling strength is increased 
\cite{Kuramoto1984}. Such a transition can 
be observed in populations of flashing fireflies, in a clapping audience, 
in coupled pendulum clocks or metronomes and in many other natural
systems \cite{Strogatz2000}.
Lately, a contrasting and more counterintuitive transition has received 
considerable interest from the nonlinear dynamics community: a population 
of identical oscillators with symmetrical coupling can split into two 
coexisting groups, one oscillating in synchrony, while the other one behaves
desynchronized. These so-called chimera states have been the subject
of several theoretical studies \cite{Kuramoto2002, Abrams2004,
  Shima2004, Sethia2008, Abrams2008, Martens2010, Omelchenko2011, 
  Omelchenko2013, Nkomo2013, Sethia2013, Schmidt2014, Sethia2014}, and 
could be realized in a number of different experimental 
systems \cite{Tinsley2012, Hagerstrom2012, Nkomo2013, Martens2013, 
  Wickramasinghe2013, Schmidt2014}; for a review, see Ref.~\cite{Panaggio2014}.

The possible importance of chimera states ranges across various 
disciplines, pertaining to phenomena such as the unihemispheric sleep 
of animals \cite{Rattenborg2000, Mathews2006, Lyamin2008}, signal 
propagation through synchronized firing in otherwise chaotic neuronal 
networks \cite{Vogels2005}, and the existence of turbulent-laminar 
patterns in Couette flow \cite{Barkley2005}. So far, chimera states 
typically show a persistent separation into coherent and incoherent 
domains, with no interchange of dynamics between the different domains. 
During unihemispheric sleep, however, when one half of the brain stays 
awake and shows desynchronized neuronal activity while the other half 
is synchronized and sleeping, the synchronization of neurons is
known to alternate between cerebral hemispheres 
\cite{Rattenborg2000, Mathews2006, Lyamin2008}.
In theoretical studies this phenomenon could only be reproduced by 
considering two man-made groups of non-identical oscillators with predefined
inter- and intra-group coupling, either autonomously \cite{Laing2012} or driven 
by a periodic external signal \cite{Ma2010}.

In contrast, in this Article we present alternating chimera states that 
spontaneously emerge in an isotropic oscillatory medium with nonlinear 
uniform global coupling, thereby tightening the connection between
chimera states and unihemispheric sleep. As the chimera states found 
so far in this system are in phase balance, and since the parameters 
are uniform and the coupling is symmetric, interchanging the incoherent 
and coherent phases again yields a solution of the underlying equations. 
A combination of local and global coupling effects then triggers the 
alternation.

\section*{Results}
The model we consider is a spatially two-dimensional system governed by a 
modified complex Ginzburg-Landau equation (MCGLE) \cite{GarciaMorales2010, Schmidt2014}:

\begin{equation}
\begin{split}
\label{eq:MCGLE}
 \partial_t W = & W + (1+i c_1)\nabla^2 W  - (1+i c_2)|W|^2 W \\
		  & - (1+i \nu)\langle W \rangle + (1+i c_2)\langle |W|^2 W \rangle\,,
\end{split}
\end{equation}

\noindent where a linear and a nonlinear global coupling term have been 
added to the standard complex Ginzburg-Landau equation (CGLE).
Here $W = W(x,y,t)$ is a complex variable describing the dynamical 
state of the system at any location $(x,y)$ at time $t$ and $\langle 
\ldots \rangle$ denotes spatial averages.  
By spatially averaging the MCGLE, one obtains a rather simple equation 
for the dynamics of $\langle W \rangle$:

\begin{equation}
 \label{eq:MCGLE_avg}
 \partial_t \langle W \rangle = -i \nu \langle W \rangle \Rightarrow \langle W \rangle = \eta \mathrm{e}^{-i \nu t}\,,
\end{equation}

\noindent displaying simple harmonic oscillations with amplitude $\eta$ and angular frequency $\nu$.
The standard CGLE is a generic model for a spatially extended system 
close to the onset of oscillations, and is considered one of the most 
important nonlinear equations in physics \cite{Aranson2002, GarciaMorales2012}.
To account for peculiar pattern formation in the oxide-layer thickness 
observed during the photoelectrodissolution of n-type silicon 
\cite{Miethe2009, Schmidt2014, Schoenleber2014}, we introduced 
the nonlinear global coupling term.
For a wide range of experimental parameters and various types of 
spatial patterns, the spatially averaged oxide-layer thickness has 
been found to display nearly harmonic oscillations 
\cite{Miethe2009, Schoenleber2014}. This is captured with the special nonlinear 
global coupling in Eq.~\eqref{eq:MCGLE}, as shown in 
Eq.~\eqref{eq:MCGLE_avg}.

\begin{figure}[h]
  \centering
  \includegraphics[width=0.98\columnwidth]{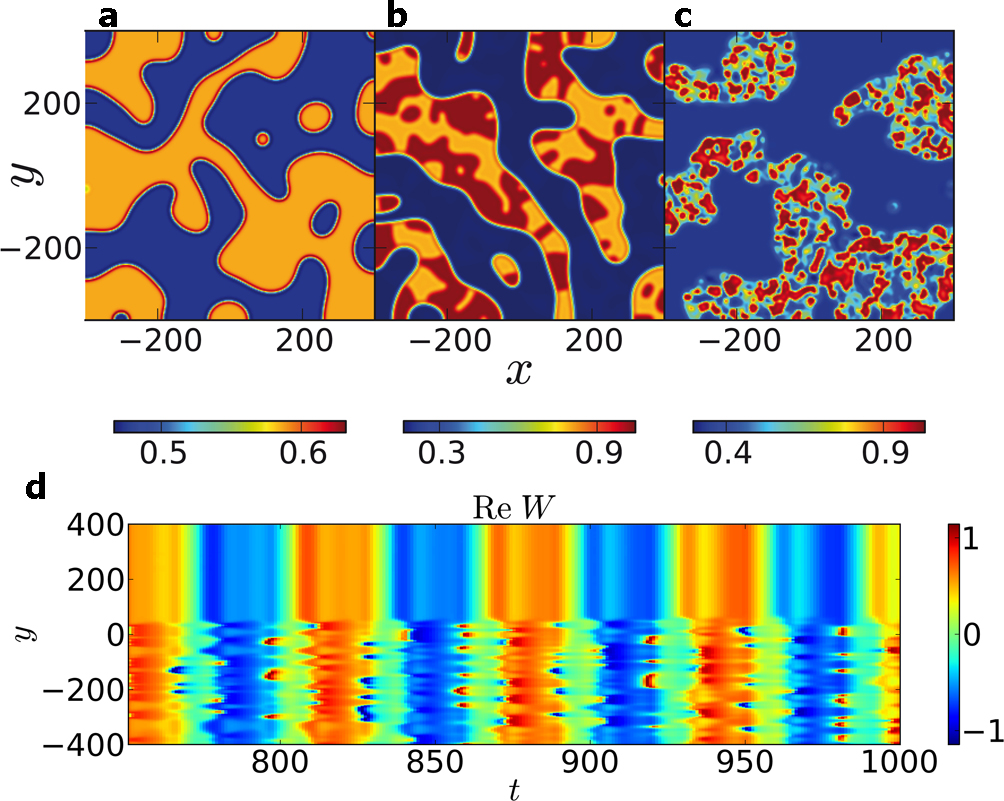}
  % W_theory.jpg: 1004x801 pixel, 300dpi, 8.50x6.78 cm, bb=0 0 241 192
  \caption{ (a)\textendash(c) Transition from two-phase clusters to a
    two-dimensional chimera state. Shown are snapshots of the real part 
    of the complex variable $W$ in (a)-(c). 
    (a) Two-phase clusters obtained for parameter $c_2 = -0.7$. Both phases are 
    homogeneous. 
    (b) Subclustering at $c_2 = -0.67$. In this case, one phase is homogeneous, 
    while the other one is split into two-phase clusters.  
    (c) Two-dimensional chimera state found for $c_2 = -0.58$. The inhomogeneous 
    phase shows strongly incoherent dynamics.
    (d) Temporal evolution of the real part of $W$ in a one-dimensional
    cut at $x=0$ in (c).
    Other parameters read: $c_1 = 0.2$, $\nu = 0.1$ and $\eta = 0.66$.
    Reprinted with permission from L.~Schmidt, K.~Sch\"{o}nleber, K.~Krischer 
  \& V.~Garc\'{\i}a-Morales, \emph{Chaos} \textbf{24,} 013102 (2014). 
  Copyright 2014, AIP Publishing LLC.} 
  % Perfectly synchronized motion coexists with asynchronous behaviour, separated by a sharp boundary.}
  \label{fig:chimera_ordinary}
\end{figure}

Numerically solving Eq.~\eqref{eq:MCGLE} for appropriate simulation parameters 
$c_1$, $c_2$, $\nu$ and $\eta$ (see methods section for details), we were able 
to reproduce different kinds of dynamics observed in the experimental 
silicon system, including a two-dimensional chimera state 
\cite{Schmidt2014}: By changing the parameter $c_2$ we observe a 
transition from two-phase clusters (Fig.~\ref{fig:chimera_ordinary}a) 
to subclustering (Fig.~\ref{fig:chimera_ordinary}b), where one of the 
two phases exhibits again two-phase clusters as a substructure. 
By further changing $c_2$ we find a two-dimensional chimera state as 
shown in Fig.~\ref{fig:chimera_ordinary}c; the spatio-temporal 
dynamics are visualized in a one-dimensional cut along $y$ in 
Fig.~\ref{fig:chimera_ordinary}d.

The chimera states previously obtained with the MCGLE exhibit 
a persistent division of the system into one coherent and 
one turbulent phase, each of which consists of one or more domains.
After their initial formation, almost no area is exchanged between 
the phases.

\begin{figure}[h]
 \centering
 \includegraphics[width=\columnwidth]{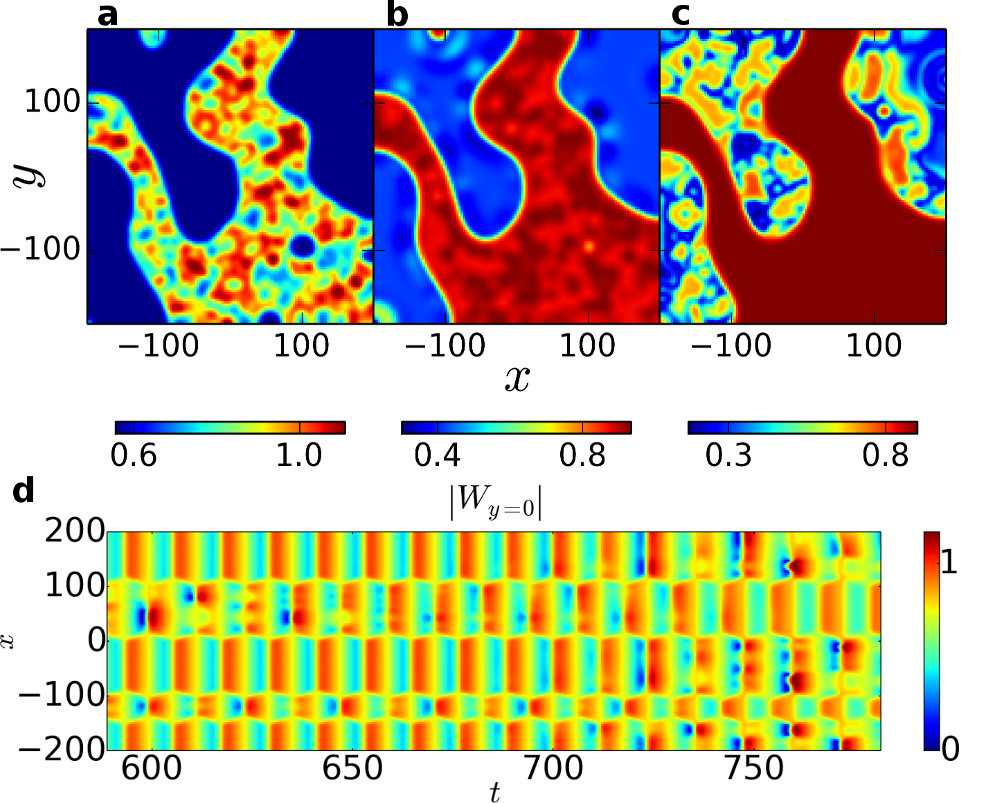}
 % W_alt_1.png: 1450x1160 pixel, 400dpi, 9.21x7.37 cm, bb=0 0 261 209
 \caption{(a) Initial coexistence of turbulence and synchrony. (b) Spread of 
turbulence to the initially synchronized phase. (c) After an interval $\Delta t	 
\approx 200$, the turbulence has moved completely from one phase to the 
other. (d) Temporal evolution of the absolute value of $W$ in a cross-section 
along the $x$-axis, covering the time interval from~(a) to~(c).}
 \label{fig:alternation1}
\end{figure}

\begin{figure}[h]
 \centering
 \includegraphics[width=\columnwidth]{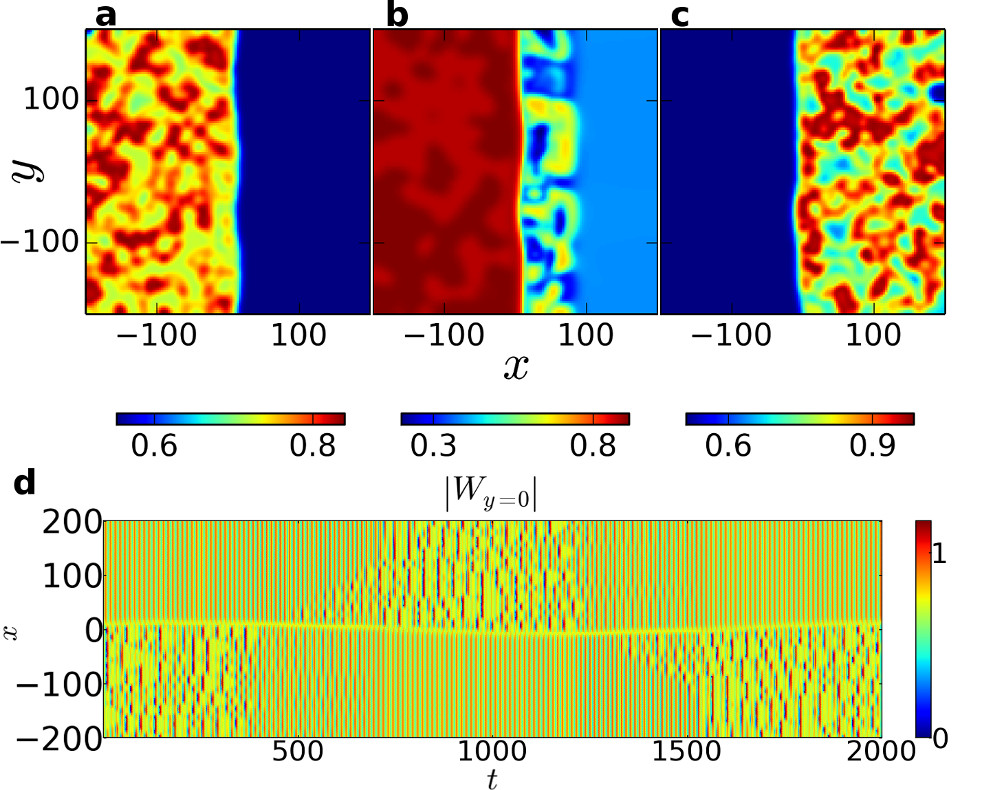}
 \caption{(a-c) Three snapshots of the system after the two-domain state has been
	reached, recorded at relative points in time $t=0, 600$ and $1000$. 
	Over the course of an interval $\Delta t~\approx 10^3$, the turbulence moves 
	completely from one phase (a) to the other (c). (d) Temporal evolution of the 
	absolute value of $W$ in a cross-section along the $x$-axis. Alternations are 
	now observed regularly at intervals of about $\Delta t \approx 10^3$.}
\label{fig:alternation_straight}
\end{figure}

Changing $c_2$ to $c_2 = -0.64$, we observe an astonishing, new kind of 
chimera state. The initial transition to a state of one synchronized 
and one turbulent phase proceeds as for the ``ordinary'' chimera states. 
However, after some time interval the dynamics in the phases interchange, 
the initially synchronized phase becoming turbulent 
while the turbulent phase becomes synchronized, as depicted in Fig.~\ref{fig:alternation1}.
Initially, these alternations do not have a characteristic timescale, but 
occur rather erratically in time.
Domains of the same phase tend to merge, thereby reducing curvature and 
length of the boundary between the phases.
Eventually (at $t=10^6$), only two domains are left, separated by a 
roughly straight boundary along one of the axes of the system. Three 
snapshots in Figs.~\ref{fig:alternation_straight}a-c visualize this
situation.
Now the alternations of turbulence and synchrony occur more regularly 
than before the two phases have properly demixed, approximately once 
every interval $\Delta t = 10^3$.
We find these approximately regular alternations to persist for at 
least $t=3.7\cdot10^7$, our maximum simulation time.
Moreover, the system was simulated with the same parameter values for 100 
slightly differing random initial conditions, always eventually leading to an 
alternation, the latest occurring at $t \approx 6 \cdot 10^4$. In contrast, the 
system was simulated with the parameter values corresponding to the 
non-alternating chimera state in Fig.~\ref{fig:chimera_ordinary}c for $t=10^6$ 
without an alternation taking place.

The two-domain state facilitates the study of the alternation 
process, which always proceeds similarly: 
First, the turbulence in the unsynchronized domain turns into a 
two-phase subclustering.
Then a spatial pattern emerges in the previously coherent domain, 
starting at the domain boundary, which acts as a nucleus for turbulence
(cf.~Figs.~\ref{fig:alternation1}b~and~\ref{fig:alternation_straight}b). 
As the incoherence spreads throughout the whole domain, the 
pattern in the other, originally turbulent domain gradually fades away, 
leaving it fully synchronized. This can be seen in 
Fig.~\ref{fig:alternation_straight}d, where the temporal development 
of a cross-section is shown. 

After the turbulence has engaged the formerly synchronized domain, the now 
turbulent domain grows further in size. This growth process is much slower 
than the preceeding spread of turbulence within the domain. Eventually, the turbulent 
domain becomes larger than the synchronized one and when a critical size is 
reached, another alternation takes place.

\begin{figure*}[ht!]
 \begin{center}
 \includegraphics[width=\textwidth]{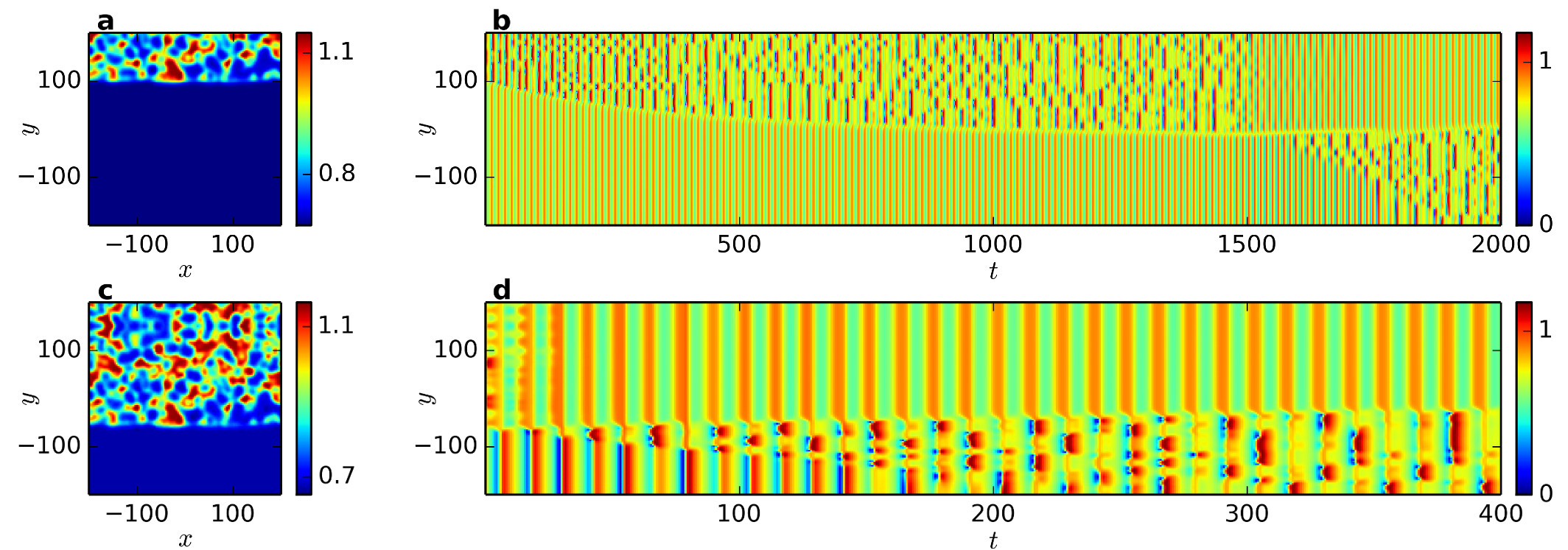}
 \end{center}
 \caption{Modified initial states not satisfying phase balance. (a) 
 Initially the turbulent domain is much smaller than the synchronized 
 domain. (c) Here, the turbulent domain is initially larger. 
 (b,d) Temporal evolution of cross-sections through the system evolving 
 from (a) and (c), respectively.}
\label{fig:domain_size_mod}
\end{figure*}

In order to validate the above proposed mechanism, we used initial 
conditions as shown in Figs.~\ref{fig:domain_size_mod}a and c, where 
either the turbulent or the synchronized domain was chosen significantly 
larger than the other one.
Starting out with turbulence covering only a small part of the 
system (Fig.~\ref{fig:domain_size_mod}a), this domain simply grows 
steadily until it covers slightly more than half the system, followed 
by an interchange of dynamics between the domains. This behavior is 
visualized in a one-dimensional cut shown in Fig.~\ref{fig:domain_size_mod}b.
Notably, the growth rate is found to be greatest at the beginning and 
to gradually decrease when approaching phase balance. This can be 
rationalized as follows: Many cluster states, including the chimera state found 
in the MCGLE, display phase balance. Thus, the phase balanced state, where 
both phases cover the same area, is a preferred state, in the sense that the 
ghost of the stable phase-balanced state is still felt. The more the 
system approaches the phase balanced state, the slower the dynamics become.

When starting the simulation from a state where most 
of the system is covered by turbulence (Fig. \ref{fig:domain_size_mod}c), 
an initial, very rapid synchronization takes place, with the 
originally turbulent domain becoming homogeneous within $\Delta t = 40$. 
This rapid synchronization is followed by a spread of turbulence throughout
the initially synchronized domain, and the same kind of steady domain growth as 
observed for the simulation where the turbulent domain is initially 
smaller. This corroborates that if the turbulent domain becomes too large, an 
alternation is triggered. Note that the growth rate of the turbulent nucleus in 
Fig.~\ref{fig:domain_size_mod}d is not symmetric in positive and 
negative $y$-direction. This could be a manifestation of the growth being 
governed by two different mechanisms. The growth process in the negative 
$y$-direction is the spreading of turbulence within the initially homogeneous 
domain. In the positive $y$-direction, the domain boundary is moving, 
at a slower pace.

For parameters corresponding to a non-alternating chimera state, 
initializing the system out of phase balance leads to the following behavior: 
an initially smaller turbulent domain grows until phase balance is 
reached, while for an initially larger turbulent domain, an 
alternation takes place at first, followed by the growth of the new, smaller 
turbulent domain up to phase balance. Thus, the difference between 
alternating and non-alternating chimeras is that in the former the phase 
balanced state is not stable and the turbulent domain grows further.

\begin{figure}[h]
 \centering
 \includegraphics[width=\columnwidth]{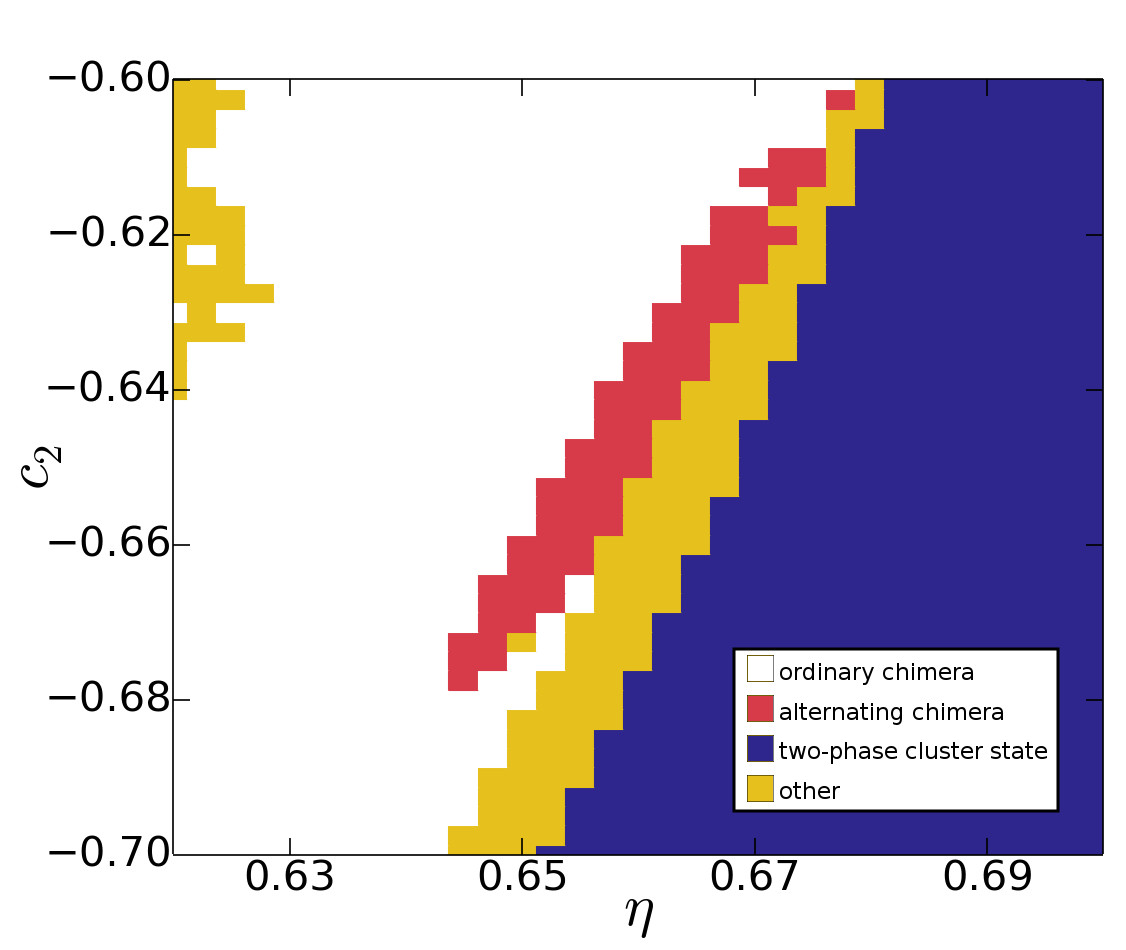}
 \caption{Phase diagram in the $c_2$ vs. $\eta$ parameter space containing 
regions of ordinary chimera 
states (white), alternating chimera states (red), two-phase cluster states 
(blue), and other dynamics (yellow), respectively. The latter group 
encompasses various different types of dynamics, including two-phase cluster 
states with subclustering, turbulence/two-phase subclustering combinations and 
two-phase states with turbulence in both phases.}
\label{fig:parameter_map}
\end{figure}

Investigating the extent of alternating chimera states in parameter space 
yields the phase diagram depicted in Fig.~\ref{fig:parameter_map}. As shown, 
they span a distinct band-like region of the $\eta-c_2$ parameter space, 
forming a part of the border between ordinary chimera states and two-phase 
cluster states. 

Moreover, when carrying out simulations for $c_2 = -0.66$ and $\eta = 0.66$, 
closer to the parameter range where two-phase subclustering was detected 
previously \cite{Schmidt2014}, the alternation of the subclustering from one of 
the phases to the other is repeatedly found as an initial transient.
However, in contrast to the alternating chimeras, this behavior has 
not yet been found to persist as long-term behavior.

\section*{Conclusion}

In summary, our simulations of a two-dimensional oscillatory medium 
governed by a complex Ginzburg-Landau equation with additional 
nonlinear global coupling, give evidence that alternating chimera 
states may spontaneously occur in isotropic oscillatory media.
The simulations suggest that alternations are the result of an 
interplay between a diffusion-driven expansion of the turbulent phase 
and a global restriction on its maximum size, as demonstrated in 
Fig.~\ref{fig:domain_size_mod}. % of the turbulent phase.
Moreover, movement of the boundary between the phases was found to 
always proceed in the direction of expansion of the turbulent phase. 
This is in accordance with earlier work on reaction-diffusion systems 
\cite{Falcke1994, Falcke1995, Hemming2002, Hemming2002a, Davidsen2005}, including 
the realistic model of catalytic CO oxidation on a $\mathrm{Pt}(110)$ 
surface \cite{Falcke1994, Falcke1995, Davidsen2005}, where the expansion 
of turbulence at the expense of synchronized domains was observed as well.

The alternating behaviour reminds of heteroclinic cycling between two 
attractors, similar to ``slow switching'' reported for two-cluster states in 
Ref.~\cite{Hansel1993, Kori2001}. However, at this state it would be premature 
to draw a conclusion about the mechanism of alternation.
% Our problem involving 
% a intrinsically turbulent region, lives in a higher-dimensional phase space, 
% and cannot be treated analytically. Hence, we lack sufficient evidence to  
% Here, however, the dynamics are 
% An alternating behaviour phenomenologically similar to the one discussed 
% here, called ``slow-switching'', has been observed earlier and clearly 
% identified as a form of heteroclinic cycling \cite{Hansel1993, Kori2001}. 
% Hovewer, due to the very high dimensionality of our system, a similarly 
% rigorous analysis of the mechanism of alternation of the alternating 
% chimera state could so far not be achieved. 

As unihemispheric sleep of animals is suggested to be a prominent example 
of chimera states emerging in biological systems \cite{Rattenborg2000, 
Mathews2006, Lyamin2008}, it is very important that the interchange of 
synchronization and incoherence between hemispheres occurring during this kind 
of sleep can be reproduced without external forcing and with identical 
oscillators. However, the current relation of chimera states and 
unihemispheric sleep has only a qualitative basis. Thus, the important 
next step in that direction would be the detailed investigation of neuronal 
dynamics during this sleep. Future research has to give answers to questions 
like what are appropriate models for neuronal oscillations and, even more 
importantly, how they are coupled.

\section*{Methods}
Simulations of Eq.\eqref{eq:MCGLE} in the main text were carried out
using a pseudospectral method, an exponential time stepping algorithm
\cite{CoxMatthews2002} and a computational timestep of $\Delta t = 0.05$. 
We used $256 \times 256$ Fourier modes, a system size of $L=400$ and no-flux 
boundary conditions. Note that the equation is dimensionless.

All simulations except those shown in Fig.~\ref{fig:domain_size_mod} were 
carried out from uniform initial conditions with superposed noise of $0.2 \%$.
Modified initial states not satisfying phase balance 
(Figs.~\ref{fig:domain_size_mod}a and c) were created by reflecting
a chimera solution about $y=\pm200$ (for a larger or smaller turbulent domain,
respectively) and choosing an appropriate section of the temporarily enlarged
system.  

Simulation parameters $c_1=0.2$, $\nu=0.1$ and $\eta=0.66$ were kept fixed for 
all simulations depicted in 
Figs.~\ref{fig:chimera_ordinary}-\ref{fig:domain_size_mod}, while $c_2$ was 
varied as described in the main text and caption of 
Fig.~\ref{fig:chimera_ordinary}.
When investigating the extent of alternating chimera states in parameter space, 
$c_1=0.2$ and $\nu=0.1$  were still left constant, while $\eta$ and $c_2$ were 
varied as shown in Fig.~\ref{fig:parameter_map}.

In order to classify the dynamics for a particular set of parameter values, 
simulations were initialized from a phase balanced state consisting of a 
homogeneous and a spatially turbulent domain, separated by a vertical boundary.
If the turbulence switched twice from one side to the other within less than 
$t=2 \cdot 10^4$, the dynamics were classified as an alternating chimera state. 
If the turbulent half remained turbulent throughout the pre-set time interval, 
while the synchronized half remained synchronized, or if they switched just 
once, the dynamics were classified as an ordinary chimera state.
Two-phase cluster states were also classified correspondingly, while all other 
dynamics were combined into a fourth group of other dynamics (see 
Fig.~\ref{fig:parameter_map}).


%\bibliography{alternating_chimera_paper}

\section*{Acknowledgments}
We gratefully acknowledge financial 
support from the \textit{Deutsche Forschungsgemeinschaft} (Grant no. 
KR1189/12-1), the \textit{Institute for Advanced Study, Technische 
Universit\"{a}t M\"{u}nchen}, funded by the German Excellence 
Initiative and the cluster of excellence \textit{Nanosystems 
Initiative Munich (NIM)}.

\section*{Additional information}

\subsection*{Author contributions}
S.W.H. carried out the simulations. L.S. and S.W.H. analyzed the data,
all authors discussed the results.
K.K. supervised the research and was responsible for the overall project
planning and directions. 
S.W.H. and L.S. wrote the paper with input from K.K. 
\\
\subsection*{Competing financial interests}
The authors declare no competing financial interests.

\end{document}